\documentclass[cameraready]{Interspeech}

\usepackage{multirow}
\usepackage{tabularx}

\usepackage{booktabs}
\title{Voice Privacy from an Attribute-based Perspective}

\author[affiliation={1}, orcid=0009-0007-4426-0321]{Mehtab Ur}{Rahman}
\author[affiliation={1,2},orcid=0000-0002-1430-4721]{Martha}{Larson}
\author[affiliation={1}, orcid=0000-0001-5395-0438]{Cristian}{Tejedor-Garcia}

\address{
    $^1$Centre for Language Studies \\
    $^2$Institute for Computing and Information Sciences\\ Radboud University, Nijmegen, The Netherlands
}

\email{ (mehtab.rahman, martha.larson, cristian.tejedorgarcia)@ru.nl}


\usepackage{comment}

\begin{document}

\maketitle

\begin{abstract}
Voice privacy approaches that preserve the anonymity of speakers modify speech in an attempt to break the link with the true identity of the speaker. Current benchmarks measure speaker protection based on signal-to-signal comparisons. In this paper, we introduce an attribute-based perspective, where we measure privacy protection in terms of comparisons between sets of speaker attributes. First, we analyze privacy impact by calculating speaker uniqueness for ground truth attributes, attributes inferred on the original speech, and attributes inferred on speech protected with standard anonymization. Next, we examine a threat scenario involving only a single utterance per speaker and calculate attack error rates. Overall, we observe that inferred attributes still present a risk despite attribute inference errors. Our research points to the importance of considering both attribute-related threats and protection mechanisms in future voice privacy research.

\end{abstract}

\section{Introduction}
\label{sec:intro}
Speech is a rich signal that conveys far more than linguistic content.
It carries a wide range of information about the speaker, including identity, physical state, health, emotional state and demographic attributes.  
Such information is useful for many applications~\cite{DEEPA2022speech,peshkova2020banking,koolagudi2012emotion}, but also constitutes sensitive personal information, leading to growing interest in voice privacy~\cite{Tom2025privacy, nautsch2019privacy}.
Until now, voice privacy research has focused on a signal-based perspective.
In other words, privacy risk is typically assessed by matching the speech signals of speakers.
A key example is the Voice Privacy Challenge (VPC)~\cite{tomashenko2020voiceprivacy, tomashenko2024voiceprivacy}, the leading benchmark in voice privacy,
other examples include~\cite{miao2025anonymization, yao2025musa}.

Such research overlooks the attribute-based perspective.
Specifically, it does not consider matching between speakers represented by profiles of categorical attributes.
Such profiles can be created by classifiers trained to infer the values of attributes from the speech signal.
Attribute inference is possible on anonymized speech; for example, the anonymization approaches in the VPC 2024~\cite{tomashenko2024voiceprivacy} are designed to preserve speaker emotion.
In general, any attribute not protected by anonymization, either intentionally or unintentionally, can be used to build a speaker attribute profile.

In this paper, we investigate the privacy of speaker attribute profiles and we point out that the attribute-based perspective on voice privacy should be considered alongside the conventional signal-oriented perspective.
The importance of the attribute-perspective on speech privacy is motivated by prior work from the wider area of data protection that has shown that categorical data can still uniquely describe individuals, even in large datasets~\cite{Montjoye2015shoppingMall, Sweeney_2002_k_anonymity, golle2006uniqueness, rocher2019re_identifications}.
In~\cite{luu2021Attribute}, the importance of speaker attributes is recognized for speaker verification as a complement to the speech signal. 
However, to our knowledge, we are the first to carry out a study of privacy protection on speaker attribute profiles and to provide a demonstration that such profiles still present a privacy risk despite attribute inference errors.

Our paper makes three main contributions:
(1) We analyze the privacy risk of speaker attribute profiles at the speaker and at the utterance level. Our analysis studies speaker uniqueness for profiles inferred from original (unprotected) speech as well as from anonymized speech. 
(2) We carry out a re-identification attack in which the attacker matches the speaker attribute profile of the target speaker that has been inferred from a single utterance (original and anonymized) to speaker attribute profiles inferred from multiple utterances of a known speaker.
(3) We release a set of annotations for four speaker attributes (gender, age, accent, and profession), aggregating and extending annotations previously released for the VoxCeleb2 dataset and enabling experimentation with attribute-based speaker profiles.\footnote{\url {https://github.com/Mehtab9/Voice-Privacy-from-an-Attribute-based-Perspective}}

\section{Related Work and Background}
\subsection{Uniqueness Analysis for Attribute-based Profiles}

An attribute-based perspective of privacy has a long history in statistical disclosure control (SDC), which studies person-level data (microdata), with a specific focus on categorical data in table form~\cite{willenborg2012disclosure, hundepool2012SDC}.
A central idea is that shared attribute values partition the dataset into equivalence classes, and privacy decreases when these classes are small.
This intuition is formalized by $k$-anonymity, where the anonymity set size $k$ is the number of records that share an attribute profile and uniqueness corresponds to $k=1$~\cite{Sweeney_2002_k_anonymity}.

The importance of uniqueness analysis is supported by regulatory and evaluation perspectives. 
Guidance related to the General Data Protection Regulation (GDPR) identifies singling out as a practical privacy risk alongside linkability and inference~\cite{EU_GDPR_2016, WP29_Opinion_05_2014}.
Although the GDPR does not formally operationalize singling out, uniqueness of attribute profiles provides a measurable interpretation of this concept.
Recent work has proposed legally grounded metrics for singling out and linkability in voice anonymization evaluation and has shown that predicate singling out risk may not correlate with speaker verification metrics such as Equal Error Rate (EER)~\cite{vauquier2025legally}. 
These considerations motivate us to use uniqueness and anonymity set sizes to assess privacy.

\subsection{Attacker Specification for the Re-identification Attack}
\label{sec:specificationattack}
The Scenario of Use Scheme~\cite{SoUS} provides a set of dimensions for the explicit specification of privacy threat situations.
Based on this scheme, the attacker scenario that forms the basis of our study of re-identification attack is specified in Table~\ref{tab:SoUS}.

\begin{table}[ht]
\scriptsize

\caption{Attacker scenario for the re-identification attack}
 \vspace{-0.1cm}
\label{tab:SoUS_Attacker}
\centering

\begin{tabularx}{\columnwidth}{p{1cm}X}
\toprule
\multirow{2}{*}{\textbf{Objective}} & The attacker aims to recover the real-world identity of a set of target (i.e., test) speakers. \\ 
\midrule
\multirow{5}{*}{\textbf{Opportunity}} & The attacker has one spoken audio utterance for each target speaker. We study two cases: the audio is original (unprotected) and the audio is protected by an anonymization algorithm. The attacker has multiple spoken utterances for a group of reference speakers that are labeled with the speaker ID.\\ \midrule
\multirow{5}{*}{\textbf{\shortstack{Additional\\Resources}}} & The attacker uses the same attribute classifiers as in the uniqueness analysis, i.e., the attacker has access to original (unprotected) speech data drawn from the same distribution as the target data that has been labeled with the four attribute classes.\\
\bottomrule
\end{tabularx}
\label{tab:SoUS}
\end{table}
The specification of the attacker scenario is based on the most recent (i.e., 2024) Voice Privacy Challenge (VPC)~\cite{tomashenko2024voiceprivacy}, which studies privacy at the utterance level. 
However, we choose to study an attack scenario more challenging than what is studied in the VPC 2024.
Specifically, for the cases involving anonymized test data, we assume that the attacker does not have access to the anonymization system and cannot anonymize the training data.

Like the VPC, our attack involves matching test speakers with reference speakers for whom the identity is known.
A key difference is that the VPC studies signal-to-signal comparisons, which allow degrees of match, and our work studies comparisons between speaker attribute profiles in terms of exact matches or mismatches.
Because we do not have degrees of match, we cannot adjust matching thresholds as needed for the Equal Error Rate used by the VPC.
Instead, we calculate an Error Rate (cf. Section~\ref{sub:att-reiden}).

\section{Experimental Setup}

\subsection{Dataset}
\label{sec:dataset}

Our research requires speech data that has been annotated with multiple speaker attributes, and we choose to build on VoxCeleb2~\cite{VoxCeleb2}, which contains speech of celebrities collected from YouTube.
Gender labels are directly available in the VoxCeleb2 metadata.
We computed age labels from the date of birth (from Wikipedia) and the recording year (from YouTube) following~\cite{Voxceleb2_age}.
Accent is approximated using nationality labels scraped from Wikipedia following~\cite{luu2021Attribute}.
Further, we generated profession labels by using information from Wikipedia and mapped them to six categories adapted from~\cite{Who_is_speaking}.

We experiment with two evaluation datasets: 
one with multiple utterances per speaker (\emph{MultiEval}), with a total of 24,588 utterances, and one with a single utterance per speaker analysis (\emph{SingleEval}), which we re-sample 10 times to control for sampling variance.
These sets contain 72 of the 118 VoxCeleb2 test speakers, which are the speakers for which all four attributes are available.
Recall from Section~\ref{sec:intro} that previous work on categorical data has shown that attribute profiles can isolate people in large scale data~\cite{Montjoye2015shoppingMall, Sweeney_2002_k_anonymity, golle2006uniqueness, rocher2019re_identifications}.
For this reason, 72 speakers are sufficient for our purpose here and to our knowledge constitute currently the largest publicly available set of speakers fully annotated with at least four attributes.
We train our attribute classifiers on the official VoxCeleb2 dev set and report classification results on the test set.
In addition to the original speech, we evaluate anonymized speech on \emph{SingleEval}, using the VPC 2024 baseline systems, McAdams (B2)~\cite{McAdams2021Patino}, STTTS (B3)~\cite{STTTS2023Sarina}, NAC (B4)~\cite{NAC2024Panariello} and ASRBN (B5)~\cite{ASRBN2023Pierre}.
Table~\ref{tab:dataset_stats} summarizes the statistics of the datasets used in our experiments.

\begin{table}[ht!]
\centering
\caption{Dataset statistics. Attributes used: Gender (2 levels), Age (3 levels), Accent (29 levels) and Profession (6 levels).
}
\resizebox{\columnwidth}{!}{%
\begin{tabular}{l p{0.7\columnwidth} r r}
\toprule
Dataset & Description & \#Speakers & \#Utterances  \\
\midrule
Classifiers train/dev set & VoxCeleb2 Dev & 5,994 & 1,092,009  \\
Classifiers evaluation set  & VoxCeleb2 Test & 118 & 36,237  \\
\midrule
\multirow{3}{*}{MultiEval} & Speaker level analysis; \newline Derived from VoxCeleb2 Test set; \newline 
mean of 341.5 utterances/speaker; \newline  ground truth, original & \multirow{3}{*}{72} & \multirow{3}{*}{24,588}  \\
\midrule
\multirow{5}{*}{SingleEval } & Utterance level analysis; \newline Derived from VoxCeleb2 Test set; \newline 10 re-samplings for each ground truth, \newline original, 4 anonymized versions & \multirow{5}{*}{72} & \multirow{5}{*}{72 $\times$ 10}  \\
\bottomrule
\end{tabular}}
\label{tab:dataset_stats}
\end{table}
\vspace{-0.4cm}

\subsection{Method for Analyzing Speaker Uniqueness}\label{sub:Uniqueness}
For our speaker uniqueness analysis, we calculate the uniqueness of speakers in our speaker set on the basis of their attribute profiles consisting of the four attributes described above: gender, age, accent, and profession.
We report uniqueness in terms of $k$, the number of speakers in the speaker set to which a given speaker is identical.
We compute the percentage of speakers in the speaker set who are unique ($k=1$) and the percentage of speakers that are below a target threshold ($k<5$), which we consider in this work as an acceptable level of $k$.  
We also report the percentage of speakers at two other thresholds ($k<3$ and $k<10$) to give a more complete picture.
Finally, we report the median $k$ over all speakers in the speaker set.

The analysis is carried out on two types of speaker profiles: first the MultiEval profiles, consisting of attributes that were inferred over multiple utterances (speaker-level profiles), and SingleEval profiles in which the attributes were inferred over a single utterance (utterance-level profiles).
The comparison reflects how privacy risk varies with the amount of speech data available.
Recall that we test 10 re-sampled SingleEval sets, so conclusions are not set-dependent.
Note that for the condition in which we use ground truth speaker attributes rather than inferred attributes, the speaker-level profiles and the utterance-level profiles are the same.

\subsection{Method for Attribute-based Re-identification Attack}\label{sub:att-reiden}
Next, we move our study of attribute-based profiles one step closer to a real-world threat scenario.
Specifically, we measure the success of a re-identification attack with the SingleEval as the target (test) data following the threat model specification in Table~\ref{tab:SoUS}. 
In the first case, the test audio is original speech (unprotected) and in the second case the test audio is protected with the four VPC 2024~\cite{tomashenko2024voiceprivacy} anonymization systems, specified in Section~\ref{sec:dataset}.
The reference data are multiple utterances per speaker from the MultiEval set, each associated with the speaker ID.
The attacker uses attribute classifiers to infer speaker attribute profiles for all test data and also for all speakers in the reference data.

The attack is performed by matching the attribute profile of each test speaker to the attribute profiles of all reference speakers.
For a given test speaker, if there is only a single reference speaker that matches, then the predicted identity of the test speaker is the identity of the matching reference speaker.
If there are multiple reference speakers that match, then the attacker makes a random selection among all matching reference speakers and the predicted identity of the test speaker is the identity of the selected reference speaker. 
We report the results of the attack in terms of the error rate, defined as the proportion of speakers that the attacker cannot identify. 


We briefly discuss the similarities and differences between this error rate and the Equal Error Rate (EER) used in the VPC. 
As previously mentioned, the EER requires the ability to control the threshold of the matching decision. 
However, in the case of attribute-based profiles, we have a binary distinction between exact match and no match and there is no threshold to adjust. 
The error rate is influenced by the process of random draw, but this contribution remains constant across all conditions. 
We report the average error rate over the 10 re-samplings of SingleEval.
Note that an attacker using partial matches would likely achieve higher attack success rates, and is relevant for future study of defenses against attribute-based attacks.

\subsection{Setup for Attribute Inference}

To train the four attribute classifiers (gender, age, accent, profession), we first extract speaker embeddings using the pre-trained ECAPA TDNN model~\cite{ECAPA_TDNN, speechbrain}, a well-established model for learning effective speaker representations. 
For each utterance, the encoder produces a 192 dimensional embedding that is used as input to lightweight attribute classifiers.
Before classification, embeddings are normalized to unit length to reduce scale variation across utterances and improve stability.
All attribute classifiers are multilayer perceptrons operating on the 192 dimensional normalized embedding.

The gender classifier is a single hidden layer MLP with ReLU activation and a single output logit for binary classification.
The age, accent, and profession classifiers are MLPs with two hidden layers using LeakyReLU activations and a final linear layer producing logits for the target number of classes.

All attribute classifiers are trained on the VoxCeleb2 development set, with 10\% of speakers per class held out for hyperparameter tuning.
Hyperparameters are selected through grid search, and the configuration that achieves the best validation performance is retained.
We retrain the final model on the full development set using the chosen hyperparameters.
At inference, attributes are predicted independently for each utterance by applying the trained classifiers and selecting the class with the highest posterior probability.
For speaker level inference, posterior probabilities from all utterances of a speaker are averaged, and the final attribute label is determined by the highest mean posterior. 
Note that we train only one set of classifiers on the original (unprotected) audio, meaning attack success will be more surprising and informative.
However, future work on defenses should also study an attacker who has access to the anonymization system used to protect speech data.

\section{Results of Analysis and Attack}

\subsection{Analysis of Speaker-Level Attribute Inference}
First, we consider the speaker-level case, in which multiple utterances are available for each speaker. 
Table~\ref{tab:attribute_classifiers} reports attribute inference performance.
In column `Speaker level' it can be seen that binary gender is inferred nearly perfectly.
Performance is substantially worse for other attributes, although it still remains above the `Weighted baseline', which is a class-weighted random classifier.
\begin{table}
\centering
\caption{Attribute classifier performance on original data. }
\vspace{-0.2cm}
\scriptsize
\begin{tabular}{lccccc}
\toprule
 & \multicolumn{1}{c}{Baselines} & \multicolumn{2}{c}{Utterance level} & \multicolumn{2}{c}{Speaker level} \\
\cmidrule(lr){2-2} \cmidrule(lr){3-4} \cmidrule(lr){5-6}
Attribute  & Weighted & Acc & F1 & Acc & F1 \\
\midrule
Gender  & 0.55 & 0.99 & 0.99  & 0.99 & 0.99 \\
Age  & 0.71 & 0.76 & 0.79 & 0.83 & 0.85 \\
Accent   & 0.39 & 0.64 & 0.66 & 0.75 & 0.73 \\
Profession   & 0.42 & 0.53 & 0.54 & 0.60 & 0.57 \\
\bottomrule
\end{tabular} %
\label{tab:attribute_classifiers}
\end{table}

\begin{table}
\centering
\caption{Speaker uniqueness with respect to speaker-level (MultiEval) attribute profiles inferred on original audio data and with respect to ground truth attribute profiles.}
\vspace{-0.2cm}
\scriptsize
\begin{tabular}{lcc}
\toprule
 & Ground truth & Inferred \\
\midrule
Unique speakers ($k = 1$) [\%] & 38.9 & 31.9 \\
$k < 3$ [\%] & 55.6 & 43.1 \\
$k < 5$ [\%] & 65.3 & 68.1 \\
$k < 10$ [\%] & 72.2 & 80.6 \\ 
\midrule
Median k & 2 & 3 \\
\bottomrule
\end{tabular}
\label{tab:kanon_speaker}
\end{table}


Table~\ref{tab:kanon_speaker} summarizes the results of the evaluation at speaker level.
The percentage of unique speakers ($k=1$) is lower when speakers are represented by inferred attributes (31.9\%) rather than ground truth attributes (38.9\%) and the median $k$ rises by one (from 2 to 3), reflecting a modest improvement in privacy.
However, considering $k=5$, which we take to be our threshold of acceptable protection, the percentage of speakers with $k<5$ rises, indicating that there are actually fewer speakers with an acceptably large group of identical speakers when speakers are represented by inferred attributes as opposed to by ground truth attributes.
Considering $k<10$ the drop is even higher.
We interpret these results to demonstrate that the noise introduced by inference leads to fewer very small groups of identical speakers (including fewer unique speakers), but also less larger groups of speakers.
In short, we do not see the errors introduced by attribute inference making a straightforward contribution to improving speaker privacy, as measured by uniqueness.

\subsection{Analysis of Utterance-Level Attribute Inference}\label{sub:UL_Inference}
Next, we consider the utterance-level case in which a single utterance is available for each speaker. 
Comparing the `Speaker level' and `Utterance level' columns of Table~\ref{tab:attribute_classifiers}, we observe that inference performance is indeed worse when only one utterance is used for prediction across all attributes, except for gender, which is still predicted nearly perfectly.

\begin{figure}[t]
  \centering
  \includegraphics[width=0.95\columnwidth]{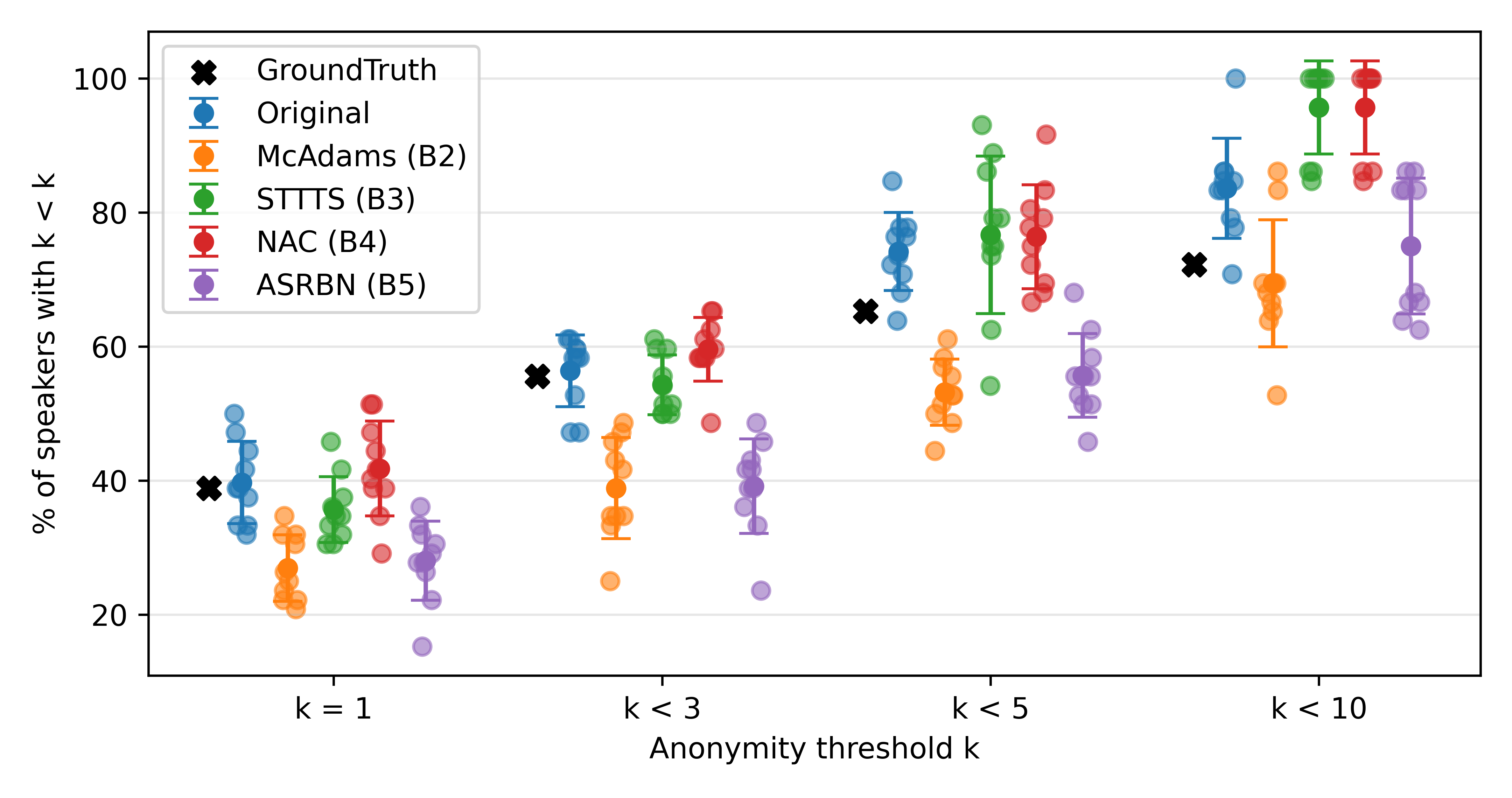}
  \vspace{-0.4cm}
  \caption{Speaker singling out in the original and anonymized single utterance per speaker setting. Each dot represents the percentage of speakers whose anonymity set size satisfies the given threshold in one independent run.}
  \label{fig:utt_per_spk_10runs_3}
\end{figure}

Figure~\ref{fig:utt_per_spk_10runs_3} reports speaker uniqueness, with the individual re-samplings of SingleEval represented as separate data points in the plot.
The black Xs indicate uniqueness percentages with speaker profiles consisting of ground truth attributes and the leftmost (blue) stacks of data points are the uniqueness percentages with speaker profiles consisting of inferred attributes from original speech. 
Considering $k=1$, the percentage of unique speakers in the case that speakers are represented with inferred attributes from the original data (Original) has a wide spread for the different resamplings.
However, on average, its percentage is the same as with the ground truth. 
The same holds for $k<3$.
For $k<5$, we again see that inference increases the percentage of speakers who fall below the acceptable threshold of $k=5$ and the same happens with $k<10$.
In short, even with only a single utterance available for inference, classifier errors are not contributing to speaker privacy in a consistent manner.

\begin{figure}[ht!]
  \centering
\includegraphics[width=0.8\columnwidth]{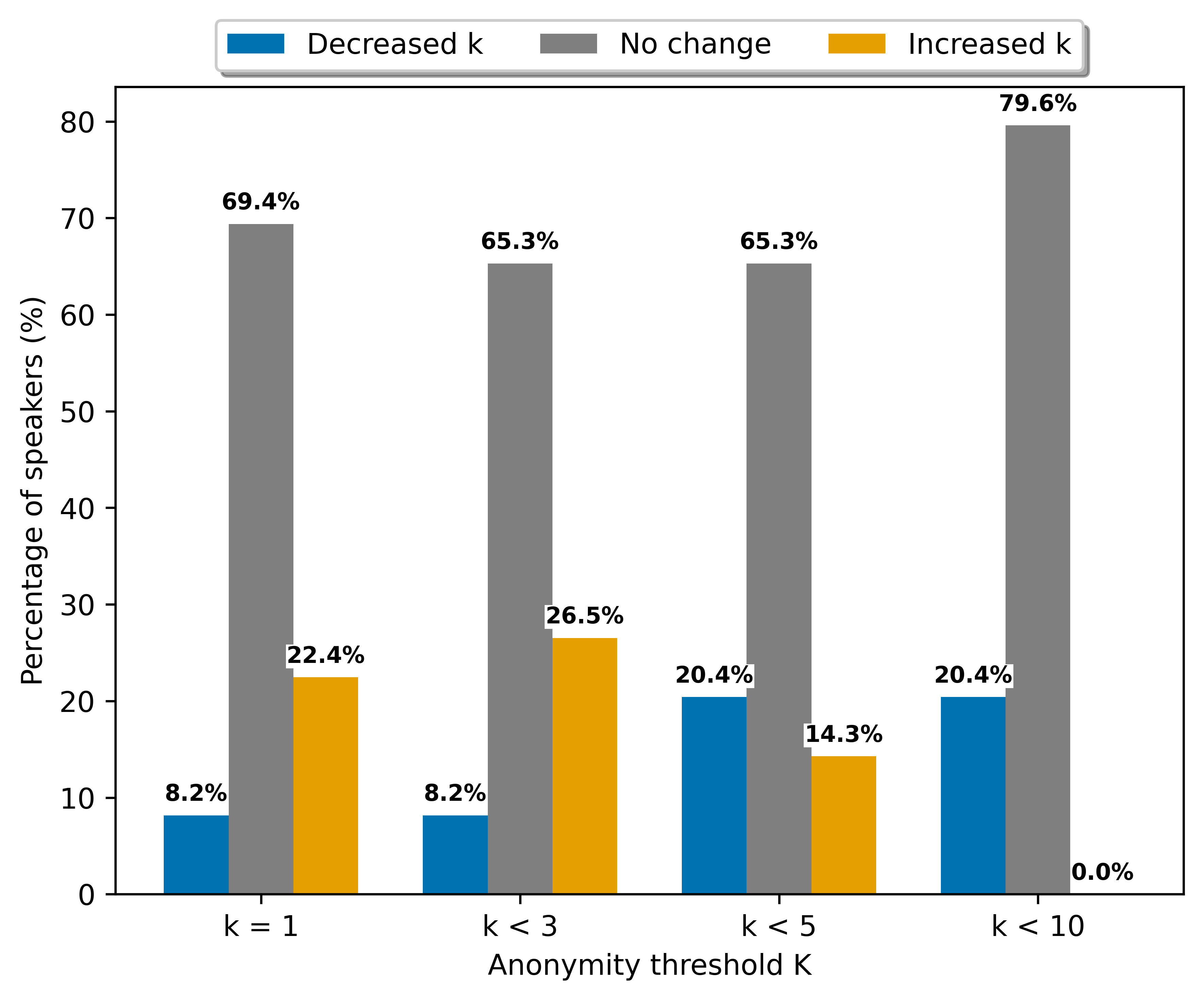}
\vspace{-0.3cm}
  \caption{Impact of attribute inference on the uniqueness across speakers.
  For different anonymity thresholds $k$, the figure shows the proportion of speakers whose anonymity set increased in size (reduced risk), decreased in size (higher risk), or remained unchanged relative to ground truth labels.}
  \label{fig:utt_per_spk_10runs}
  \vspace{-0.2cm}
\end{figure}

To gain deeper understanding, we consider what changes at the level of individual speakers. Figure~\ref{fig:utt_per_spk_10runs} shows that the change in uniqueness does not impact all speakers equally.
The bars in this graph show the percentage of speakers at each $k$ that becomes lower (left), meaning their situation is worse with respect to their $k$-level, stays the same (middle), and becomes higher (right side), meaning their privacy situation has become better. 
At the $k<10$ level, inference makes the situation worse for 20.4\% of the speakers and improves the situation for no one. 

\subsection{Analysis of Inference on Anonymized Utterances}
We now turn to the case in which a single utterance is available per speaker, and that utterance has been anonymized.
It is important to note that, in general, anonymization techniques are developed to protect against de-anonymization, but not specifically designed to also offer full protection against attribute inference.
The accuracy levels of inference on anonymized speech are quite low, as can be seen in Table~\ref{tab:attribute_methods}.
Recall that the attribute inference classifier is trained on unprotected speech.

\begin{table}
\centering
\caption{Attribute classification accuracy (mean $\pm$ std) for different anonymization methods}
\scriptsize
\begin{tabular}{lcccc}
\toprule
Attribute & McAdams & STTTS & NAC & ASRBN \\
\midrule
Gender      & 0.80 $\pm$ 0.03 & 0.47 $\pm$ 0.05 & 0.67 $\pm$ 0.04 & 0.53 $\pm$ 0.03 \\
Age         & 0.56 $\pm$ 0.05 & 0.31 $\pm$ 0.05 & 0.65 $\pm$ 0.05 & 0.36 $\pm$ 0.07 \\
Accent      & 0.52 $\pm$ 0.05 & 0.38 $\pm$ 0.05 & 0.38 $\pm$ 0.05 & 0.49 $\pm$ 0.04 \\
Profession  & 0.57 $\pm$ 0.03 & 0.26 $\pm$ 0.06 & 0.38 $\pm$ 0.05 & 0.39 $\pm$ 0.05 \\
\bottomrule
\end{tabular}%
\label{tab:attribute_methods}
\end{table}

In Figure~\ref{fig:utt_per_spk_10runs_3}, we see that there is no large difference between the cases in which attributes are inferred from original speech and from speech anonymized with NAC or STTTS. 
McAdams and ASRBN provide protection by lowering the number of speakers with $k<5$, but none of the approaches differs substantially from the ground truth in the case $k<10$.
In short, analysis reveals that the attribute inference error on anonymized speech is also not consistently contributing to improved privacy. 

\subsection{Re-Identification Attack}
\label{sec:ReIdentificationAttack}


The re-identification attack compares target speaker profiles to reference profiles.
Table~\ref{tab:attack_error_combined} shows the attack performance when the target speaker profile is inferred from original speech.
For original target speech, the attack error rate is lower when the reference profile is inferred from original speech rather than when the ground truth profile is used. 
We relate this to a correlation in the errors of the attribute classifiers.

Table~\ref{tab:attack_error_combined} further shows that attacks based on attribute profiles inferred from anonymized speech can, in some cases, achieve error rates comparable to, or even lower than, those observed for original speech.
A priori, it might be tempting to conclude that noise introduced by imperfect inference of attributes helps to protect privacy.
However, these results suggest that correlated inference errors between target and reference profiles can instead reduce privacy.
If the classifiers were to predict the majority class for each attribute, all speakers would be identical and the attack error rate would be $1-1/72=0.986$.

Note that in Table~\ref{tab:attack_error_combined}, the target speech is anonymized, while the reference speech remains unprotected. Therefore, these results are not directly comparable to the speaker uniqueness results in Section~\ref{sub:UL_Inference}.

\begin{table}[ht]
\centering
\caption{Attack error rate. Two types of speaker profiles used as reference. Inferred reference profiles are at speaker level.}
\label{tab:attack_error_combined}
\scriptsize
\begin{tabular}{lcc}
\toprule
Target speaker profile
& Reference:  
& Reference:  \\
inferred on 
& ground truth profile 
& profile inferred from original \\
\midrule

Original (unprotected) 
& 0.72 $\pm$ 0.04 
& 0.67 $\pm$ 0.03 \\
\midrule

McAdams (B2)   
& 0.76 $\pm$ 0.04 
& 0.78 $\pm$ 0.04 \\

STTTS (B3)     
& 0.62 $\pm$ 0.05 
& 0.82 $\pm$ 0.04 \\

NAC (B4)     
& 0.70 $\pm$ 0.06 
& 0.71 $\pm$ 0.05 \\

ASRBN (B5)  
& 0.58 $\pm$ 0.05 
& 0.82 $\pm$ 0.04 \\

\bottomrule
\end{tabular}
\vspace{-0.3cm}
\end{table}

\section{Conclusion and Outlook}
We have taken a first look at voice privacy from an attribute-based perspective, studying speaker attribute profiles comprised of attributes inferred from spoken audio. 
We have shown that although the performance of attribute inference classifiers may be low, misclassifications do not always provide additional privacy protection.
Specifically, misclassifications cause some speakers to become more unique (Figure~\ref{fig:utt_per_spk_10runs}) and can lower the number of speakers who enjoy a sufficiently large number of identical speakers (here taken to be $k\ge5$) (Table~\ref{tab:kanon_speaker} and Figure~\ref{fig:utt_per_spk_10runs_3}).
Further, the very low attribute classification performance on anonymized data does not always contribute to lowering the uniqueness of speakers, as shown by our uniqueness analysis (Figure~\ref{fig:utt_per_spk_10runs_3}).
Also, it does not always contribute to substantially raising the speaker re-identification error rate as shown by our re-identification attack  (Table~\ref{tab:attack_error_combined}).

Future work should examine the patterns of classifier mistakes, since attribute classifiers that are consistent in their inaccurate predictions of the attributes for a given speaker can raise uniqueness and lower attack error.
Looking forward, our work has laid the groundwork for future study of privacy from an attribute-based perspective as well as the development of defenses against attribute-based attacks.

 \section{Acknowledgments}

This publication is part of the project Responsible AI for Voice Diagnostics (RAIVD) with file number NGF.1607.22.013 of the research program NGF AiNed Fellowship Grants, which is financed by the Dutch Research Council (NWO).

 \ifcameraready

 \else
      Anonymous content.
 \fi

\section{Generative AI Use Disclosure}
Generative AI tools were used for language editing and polishing, including grammar and phrasing. All scientific content, experimental design, analyses, results, and conclusions were developed, verified, and approved by the authors. The authors take full responsibility for the content of this paper, and no generative AI tool is listed as a co-author.

\bibliographystyle{IEEEtran}
\bibliography{mybib}

\end{document}